\documentclass[%
 reprint,
 amsmath,amssymb,
 aps,
]{revtex4-2}

\usepackage{graphicx}
\usepackage{dcolumn}
\usepackage{bm}
\usepackage{xcolor}
\usepackage{etoolbox}
\usepackage[final]{changes}

\makeatletter
\undef\c@Changes@AddCount%
\@namedef{Changes@AuthorColor}{red}
\colorlet{Changes@Color}{red}
\makeatother

\newcommand{\Dc}{D_\mathrm{c}}
\newcommand{\Dm}{D_\mathrm{m}}
\newcommand{\xmi}{x_{i}}
\newcommand{\xmj}{x_{j}}
\newcommand{\xc}{x_\mathrm{c}}
\newcommand{\dotxmi}{\dot{x}_{i}}
\newcommand{\dotxc}{\dot{x}_\mathrm{c}}

\newcommand{\ri}{\Delta \xmi}
\newcommand{\rj}{\Delta x_j}
\newcommand{\force}{f_\mathrm{chem}}
\newcommand{\fano}{\phi}
\newcommand{\CV}{\theta}

\newcommand{\PMi}{P_{\to \mathrm{M}_i}}
\newcommand{\PMC}{P_{\mathrm{M}_i\to \mathrm{C}}}
\newcommand{\QC}{\dot{Q}_{\to \mathrm{C}}}
\newcommand{\QMi}{\dot{Q}_{\to \mathrm{M}_i}}
\newcommand{\Pchem}{P_\mathrm{chem}}
\newcommand{\etaM}{\eta_\mathrm{M}}
\newcommand{\etaS}{\eta_\mathrm{S}}

\setaddedmarkup{\textcolor{red}{#1}}
\setdeletedmarkup{\textcolor{black}{\sout{#1}}}

\begin{document}

\preprint{APS/123-QED}

\title{Performance scaling and trade-offs for collective motor-driven transport}

\author{Matthew P.\ Leighton}
\email{matthew\_leighton@sfu.ca}
\author{David A.\ Sivak}%
 \email{dsivak@sfu.ca}
\affiliation{Department of Physics, Simon Fraser University, Burnaby, British Columbia, V5A 1S6, Canada.}%

\date{\today}

\begin{abstract}
Motor-driven intracellular transport of organelles, vesicles, and other molecular cargo is a highly collective process. An individual cargo is often pulled by a team of transport motors, with numbers ranging from only a few to several hundred. We explore the behavior of these systems using a stochastic model for transport of molecular cargo by an arbitrary number $N$ of motors obeying linear Langevin dynamics, finding analytic solutions for the $N$-dependence of the velocity, precision of forward progress, energy flows between different system components, and efficiency. In two opposing regimes, we show that these properties obey simple scaling laws with $N$. Finally, we explore trade-offs between performance metrics as $N$ is varied, providing insight into how different numbers of motors might be well-matched to distinct contexts where different performance metrics are prioritized.
\end{abstract}

\maketitle

\section{Introduction}
Living organisms are fundamentally out-of-equilibrium physical systems~\cite{schrodinger1944life} characterized by sustained spatial inhomogeneities. To maintain this nonequilibrium state, organisms must constantly consume energy and transport material at several different lengthscales, including within individual cells. Intracellular transport is achieved using a plethora of different methods including passive and active diffusion, advection, ion pumps, and motor-driven transport~\cite{mogre2020getting}. These transport processes are characterized by unavoidable stochasticity as well as the overdamped motion inherent to the low-Reynolds-number regime these systems inhabit~\cite{purcell1977life}.

Motor proteins (also referred to as `transport motors' or just `motors') are integral components of eukaryotic cells, with a wide range of functions including transport of large macromolecular cargo over significant distances~\cite{hancock2008intracellular}. These motors transduce chemical energy, generally in the form of ATP, into net mechanical motion in a preferred direction~\cite{brown2019theory}. Indeed, motor proteins can be thought of and modeled as nanoscale thermodynamic engines, whose behavior is characterized by stochastic mechanical and chemical dynamics~\cite{seifert2012stochastic}. Individual transport motors can reach speeds as high as $\sim$$8\,\mu\mathrm{m/s}$~\cite{purcell2011nucleotide} and make forward progress while pulling against forces on the order of $\sim$$ 6\,\mathrm{pN}$~\cite{fallesen2011force}. Particularly well-characterized examples include kinesin and dynein motors pulling vesicles along microtubules~\cite{encalada2011stable}, and myosin motors pulling on actin filaments to contract muscle tissue~\cite{cooke1997actomyosin}.

Motor proteins within cells often work collectively to transport large organelles such as mitochondria~\cite{hancock2008intracellular,leopold1992association}, or even chromosomes during mitosis and meiosis~\cite{hatsumi1992mutants}. Experimental determination of the number $N$ of motors attached to a given cargo is generally challenging; nonetheless, recent studies have successfully measured $N$ by identifying discrete peaks in a distinctly multimodal velocity distribution~\cite{shtridelman2008force,shtridelman2009vivo} for small numbers of motors, or using more complex techniques such as quantitative immunoblots and immunoelectron microscopy~\cite{leopold1992association}. Experimental investigations both \textit{in vivo} and \textit{in vitro} have found widely varying numbers $N$ of motors coupled to a single cargo. In some cases, only a single motor~\cite{svoboda1994force} or a few motors~\cite{shtridelman2008force} per cargo is observed, but experiments have observed as many as $200$ motors bound to large organelles~\cite{leopold1992association}. Likewise, in actomyosin filaments in muscle tissue, on the order of $100$ motors are attached to each actin filament~\cite{rastogi2016maximum}. Collective-transport systems can also be engineered \textit{in vitro}~\cite{delrosso2017exploiting}; in this setting the number of motors can be controlled more precisely, for example using DNA scaffolds~\cite{furuta2013measuring,derr2012tug}.

Simple phenomenological models for transport-motor dynamics (such as the classical linear force-velocity relationship~\cite{svoboda1994force,hunt1994force}) has been extended to multiple motors, for example by assuming equal load-sharing~\cite{shtridelman2008force,shtridelman2009vivo}. These models provide a good first approximation to the dynamics of multi-motor systems, and can be extended to include, for example, motor binding and unbinding kinetics~\cite{klumpp2005cooperative,kunwar2010robust}. These types of models assume that the motors pull against a constant force, rather than the stochastic frictional drag that would occur for a loosely coupled diffusive cargo. However, analysis of transport by single motors has shown that pulling a diffusing cargo and pulling against a constant force lead to qualitatively different transport behavior~\cite{zimmermann2015effective,brown2019pulling}. Researchers have proposed and explored several dynamical models for transport of diffusive cargo by multiple motors ~\cite{korn2009stochastic,mckinley2012asymptotic,li2012critical,bhat2016transport,bhat2017stall,arpaug2014transport}. These approaches rely primarily on numerical simulation, and as such are limited by computational resources to exploring systems with relatively small numbers of motors.

In studying intracellular transport, an important goal is to understand how systems can be tuned to improve performance. Relevant performance metrics vary based on the context, but may include cargo velocity, rate of chemical energy consumption, transport efficiency, and precision~\cite{brown2019theory}. The dependence of these and other performance metrics on the number of motors is of clear interest, and has not yet been systematically investigated.

In this article we introduce a simple, thermodynamically consistent, stochastic model for the collective transport of diffusive molecular cargo by an arbitrary number $N$ of motors. This model has a key advantage over other recent theoretical and computational approaches: we can solve it analytically for arbitrary $N$, allowing us to explore system behavior over many orders of magnitude of motor numbers. We derive $N$-dependent expressions for several performance metrics, and explicitly calculate all thermodynamic energy flows between different system components and thermal and chemical reservoirs. This allows us to derive simple analytic expressions for efficiency both of the whole system and of individual motors. In two opposing regimes we identify simple scaling laws that characterize the $N$-dependence of these properties. Finally, we derive fundamental trade-offs among these performance metrics, thereby pointing to design principles for collective motor-driven transport.

\section{Model and Theory}
\begin{figure}[h]
\includegraphics[width=\columnwidth]{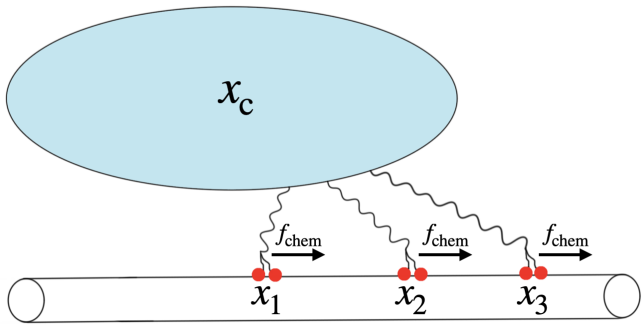}
\caption{\label{fig:diagram} Collective-transport system comprising a single diffusive cargo coupled to $N$ (here $N=3$) motor proteins moving along a linear substrate. $\xc$ indicates the cargo position, and $\xmi$ the position of the $i$th motor. Each motor experiences chemical driving force $f_\mathrm{chem}$.}
\end{figure}

We consider a diffusive cargo coupled to $N$ identical transport motors, with motion resolved in one dimension. Each motor interacts only with the cargo via a molecular linker, and is characterized by a mechanochemical cycle through which it transduces chemical power into directed forward motion. The cargo undergoes Brownian motion subject to coupling forces from each motor via the respective linker. Figure~\ref{fig:diagram} illustrates the system.

We focus on the limiting regime in which the time evolution of the system is independent of the initial conditions, which we call its steady state. Mathematically, the relevant limit is that the time greatly exceeds the system's longest relaxation time ($t\gg\tau_\mathrm{relax}$). In our general discussion, we assume that this limit exists, and that in the steady state, system properties such as the velocity, efficiency, energy flows and entropy production all have well-defined constant average values. The specific model we introduce below satisfies these assumptions.

\subsection{Model}\label{specific_model}
We model the cargo and motors as overdamped Brownian particles diffusing (with respective diffusivities $\Dc$ and $\Dm$) in a potential landscape. $\xc$ and $\xmi$ describe the positions of the cargo and $i$th motor. For mathematical simplicity we treat the cargo as a single point, but our model describes equally well (through a linear change of variables) motors attached at different points to a rigid cargo. The system is isothermal and in contact with a thermal reservoir at inverse temperature $\beta\equiv (k_BT)^{-1}$. We assume that both cargo and motor dynamics satisfy the fluctuation-dissipation relation: the friction coefficients for the cargo ($\zeta_\mathrm{c}$) and motors ($\zeta_\mathrm{m}$) are related to their respective diffusivities by $\beta \zeta_\mathrm{c} \Dc=1 = \beta \zeta_\mathrm{m}\Dm$.

Each motor is subject to chemical driving due to nonequilibrium environmental concentrations of molecules, often ATP, ADP, and phosphate, and further experiences an underlying periodic energy landscape due to interactions with the substrate it walks along. To simplify the analysis, we assume that the scale of this energy landscape (the heights of the barriers separating meta-stable states) are small compared to the magnitude of the chemical driving.
This leads to the motors experiencing a flat, downward-sloping energy landscape, which can be thought of as a constant force $\force$ propelling each motor in a preferred direction. We further assume tight coupling between chemical-energy consumption and forward motor motion.

We model each linker coupling one motor to the cargo as a Hookean spring with zero rest length, dominated by the along-filament displacement, thus with interaction potential $U_i(\xc,\xmi) = \frac{1}{2} \kappa (\xmi-\xc)^2$. This is a common assumption in modeling approaches~\cite{bhat2016transport,brown2019pulling} and experimentally well-supported for kinesin linkers~\cite{kawaguchi2003equilibrium}.

This model system dynamically evolves according to $N+1$ coupled Langevin equations,
\begin{subequations}\label{multiplemotorlangevin}
\begin{align}
\dotxc & = \beta \Dc \kappa \sum_{i=1}^N \left(\xmi - \xc\right) + \eta_\mathrm{c},\label{eq:cargoMotion}\\
\dotxmi & = \beta \Dm\left[\force - \kappa (\xmi - \xc)\right] + \eta_{i},\; i = 1,\ldots,N. 
\label{eq:motorMotion}
\end{align}
\end{subequations}
Here $\eta_{\rm c}(t)$ and $\eta_i(t)$ are independent Gaussian noises with means and variances
\begin{subequations}
\begin{align}
\langle \eta_\mathrm{c}(t)\rangle & = 0,\\
\langle \eta_{i}(t)  \rangle & = 0 \\
\langle \eta_\mathrm{c}(t)\eta_\mathrm{c}(t')\rangle & = 2\Dc\delta(t-t'),\\
\langle \eta_{i}(t)\eta_{j}(t')\rangle & = 2\Dm\delta_{ij}\delta(t-t'),\\
\langle \eta_\mathrm{c}(t)\eta_i(t')\rangle & = 0.
\end{align} 
\end{subequations}
Here $\delta_{ij}$ is the Kronecker delta function and $\delta(t-t')$ the Dirac delta function. 

The single-motor dynamics~\eqref{eq:motorMotion} produce average motion equivalent to the linear force-velocity relation typically observed experimentally for kinesin motors under constant forces less than their stall force~\cite{svoboda1994force,hunt1994force} (where most of our analysis takes place),
\begin{equation}\label{force_velocity}
\langle v\rangle = v_\mathrm{max}\left(1 - \frac{f}{f_\mathrm{s}}\right),
\end{equation}
for stall force $f_\mathrm{s} = \force$, maximum velocity $v_\mathrm{max} = \beta \Dm\force$, and force $f=\kappa\langle \xmi-\xc\rangle$ acting on the motor.

\subsection{Thermodynamics}
Our model is thermodynamically consistent, so we use the tools of stochastic thermodynamics~\cite{seifert2012stochastic} to analyze the energy flows between various system components. Each subsystem (cargo or motor) exchanges heat with a thermal reservoir at inverse temperature $\beta$, and each motor exchanges chemical energy with reservoirs characterized by constant chemical potentials. The $i$th motor exchanges energy with the cargo through their interaction potential $U_i$. 

The $i$th motor consumes average chemical power 
\begin{equation}\label{motor_power_eq}
\PMi = \left\langle \dotxmi \circ f_\mathrm{chem}\right\rangle
\end{equation}
and transmits average power 
\begin{equation}\label{power_transferred}
\PMC = \left\langle \dotxmi \circ \frac{\partial U_i}{\partial \xmi}\right\rangle
\end{equation}
to the cargo through the linker. Equation~\eqref{motor_power_eq} implicitly assumes each tightly couples its chemical and mechanical degrees of freedom, consistent with experimental findings for kinesin motors~\cite{schnitzer1997kinesin,visscher1999single}. The respective average rates of heat flow from the thermal reservoir into the cargo and $i$th motor are
\begin{subequations}\label{heat_flows}
\begin{align}
\QC & = \left\langle \dotxc \circ \sum_{i=1}^N \frac{\partial U_i}{\partial \xc}\right\rangle,\\
\QMi & = \left\langle \dotxmi \circ \left[\frac{\partial U_i}{\partial \xmi} - f_\mathrm{chem}\right]\right\rangle.
\end{align}
\end{subequations}
Here, angled brackets denote ensemble averages over stochastic fluctuations, and the product (indicated by the symbol $``\circ"$) between $\dotxc$ or $\dotxmi$ and other quantities is interpreted using the Stratonovich rule so that the ensemble averages can be evaluated using the methods outlined in \cite{seifert2012stochastic}. Transport systems are highly processive, so we focus on average energy flows, ignoring higher moments that are less salient at long durations.  

At steady state, the average internal energy of each subsystem is constant, giving subsystem-specific first laws
\begin{subequations}\label{first_law}
\begin{align}
0 & = \sum_{i=1}^N\PMC + \QC,\\
0 & = \PMi - \PMC + \QMi,\; \forall i\in 1,\ldots,N.
\end{align}
\end{subequations}
The second law bounds the total entropy production rate $\dot{\Sigma}$ (here equaling the total chemical power $\Pchem = \sum_i \PMi$) and hence the heat flows~\cite{horowitz2015multipartite}:
\begin{subequations}\label{second_law_heat}
\begin{align}
0 \leq \dot{\Sigma} & = \Pchem\\
& =-\QC - \sum_{i=1}^N\QMi.
\end{align}
\end{subequations}

\section{Results}

\subsection{Solution}\label{Solution_timescale}

Equations~\eqref{multiplemotorlangevin} constitute a linear system of coupled Langevin equations, and as such are in general analytically solvable, with solution a multivariate Gaussian. Thus it suffices to solve for the mean vector and covariance matrix of the whole system, the components of which satisfy a set of coupled linear ordinary differential equations~\cite[Section 3.2]{risken1996fokker}. Since the $N$ motors dynamically evolve according to identical stochastic equations~\eqref{eq:motorMotion}, their marginal position distributions are identical. As a result, there are only two unique means ($\langle \xmi\rangle$ and $\langle \xc\rangle$) and four unique covariances ($\mathrm{Cov}(\xc,\xc), \mathrm{Cov}(\xc,\xmi), \mathrm{Cov}(\xmi,\xmi)$, and $\mathrm{Cov}(\xmi,\xmj$), all of which vary with time. This symmetry permits exact solution without specifying $N$. Appendix~\ref{ana_mu_cov} details the unwieldy analytic expressions.

The distributions of $\{\xmi(t)\}_{i=1}^N$ and $\xc(t)$ are time-dependent, so we change to a set of $N$ variables, $\ri(t) = \xmi(t) - \xc(t)$, that at steady state converge to a time-independent joint distribution, a multivariate Gaussian with means and covariances
\begin{subequations}\label{eq:r_dist}
\begin{align}
\langle \ri\rangle & = \frac{\force}{\kappa} \left(1 + N\frac{\Dc}{\Dm}\right)^{-1}, \label{eq:mean_r}\\
\text{Cov}(\ri,\rj) & = \frac{1}{\beta \kappa} \delta_{ij}.
\end{align}
\end{subequations}
This time-independent distribution is sufficient to compute many steady-state properties of interest.

For our model we find that the off-diagonal entries of the stationary covariance matrix, $\text{Cov}(\ri,\rj)$ for $i\neq j$, are zero: fluctuations in the relative position of one motor are uncorrelated with the relative positions of the other motors. We do not expect this particular result to generalize; for example, collective-transport models with discrete motor motion~\cite{bhat2016transport} have found non-zero off-diagonal covariances at small $N$. Regardless, the results presented below depend only on diagonal terms and are independent of off-diagonal covariances.

The system relaxation time is $\tau_\mathrm{relax} = \left[ \beta\kappa (\Dm + N\Dc)\right]^{-1}$. Appendix~\ref{params} provides parameter estimates indicating that this is at most $\sim$ $0.02$ s for kinesin pulling molecular cargo, and generally at least an order of magnitude smaller. This corresponds to a distance of at most $40\,\mathrm{nm}$ for kinesin motors at maximum velocity. Given the short distance over which relaxation to steady state occurs, and the high processivity of motor-driven transport systems (kinesin can travel up to several micrometers before detaching~\cite{toprak2009kinesin}), we exclusively focus on the steady state.

The dimensionless parameter combination $N\Dc/\Dm$ appears in Eq.~\eqref{eq:mean_r} and in many of the results shown later on, constituting a key quantity for understanding the system behavior. For intracellular transport of vesicles or organelles the diffusivity ratio $\Dc/\Dm$ is typically $\sim$ $ 10^{-3} - 10^0$, depending on the size of the cargo (see Appendix~\ref{params}). Since $N$ can range from one to several hundred, we focus our explorations on the range $N\Dc/\Dm\in[10^{-3},10^3]$.

\subsection{Scaling behavior}

\subsubsection{Dynamical properties}\label{sec:dynamical}
\begin{figure}[t]
\includegraphics[width=\columnwidth]{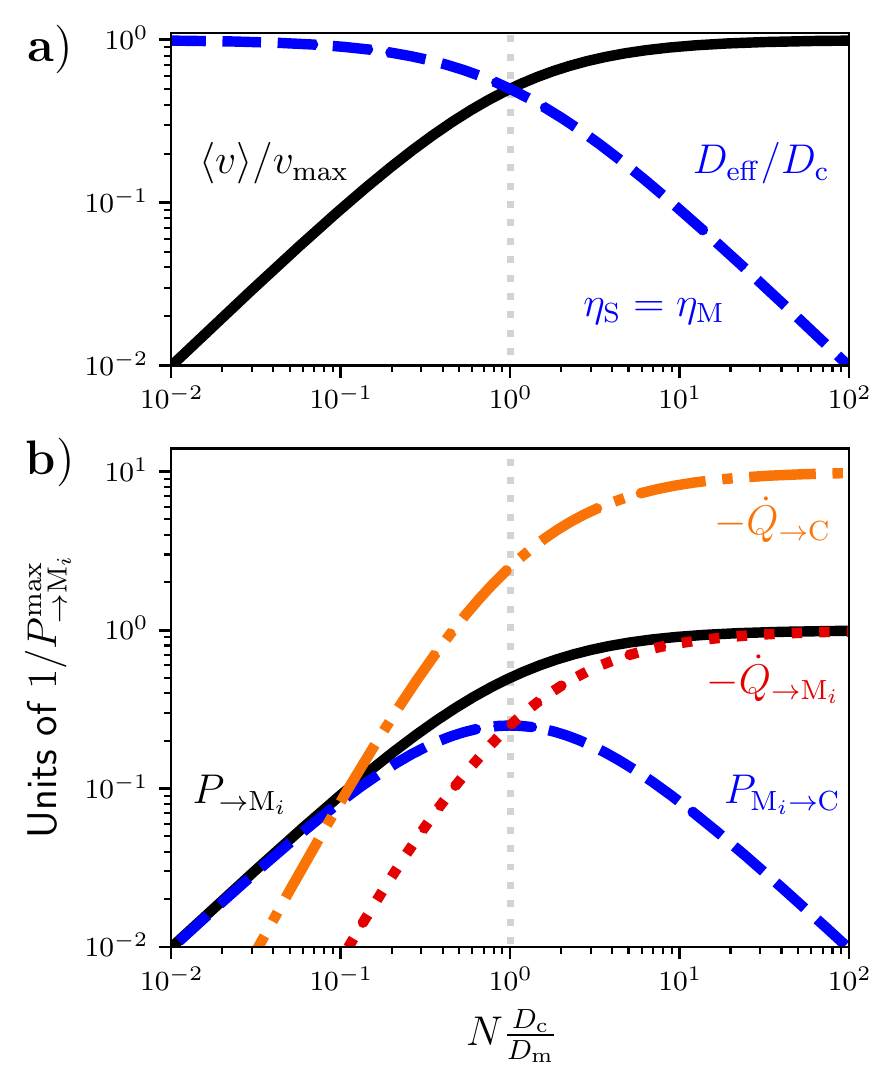}
\caption{\label{fig:properties} \textbf{a)} Scaled velocity $\langle v\rangle/v_\mathrm{max}$ (Eq.~\eqref{velocityeq}) and effective diffusivity $D_\mathrm{eff}/D_\mathrm{c}$ (Eq.~\eqref{eq:D_eff}) (equal to the efficiency $\eta\equiv\etaM=\etaS$ (Eq.~\eqref{effic_eq})) as functions of the number $N$ of motors scaled by the diffusivity ratio $\Dm/\Dc$. \textbf{b)} Ensemble-averaged energy flows $\PMi$ (Eq.~\eqref{eqn:P_to_M}), $\PMC$ (Eq.~\eqref{eqn:P_M_to_C}), $\QMi$ (Eq.~\eqref{eqn:Q_to_M}), and $\QC$ (Eq.~\eqref{eqn:Q_to_C}) as functions of $N$ scaled by $\Dc/\Dm$. All energy flows are scaled by the chemical power consumption $\PMi^\mathrm{max} = \force v_\mathrm{max}$ of a single motor at maximum velocity. For $\QC$ we use $\Dc/\Dm=1/10$; all other energy flows depend solely on the quantity $N\Dc/\Dm$.}
\end{figure}

At steady state, the cargo and each motor have equal velocity, $\langle v_{\rm c}\rangle \equiv \lim_{t\to\infty}\langle x_{\rm c}(t) - x_{\rm c}(0)\rangle/t = \lim_{t\to\infty}\langle \xmi(t) - \xmi(0)\rangle/t \equiv \langle v_{\rm m}\rangle \equiv \langle v \rangle$. Evaluating this limit yields a simple expression for the $N$-dependent mean system velocity:
\begin{equation}\label{velocityeq} 
\langle v\rangle = v_\mathrm{max}\left(1+ \frac{\Dm}{N\Dc}\right)^{-1}. 
\end{equation}
For $N\ll \Dm/\Dc$, the mean velocity grows linearly with $N$, so adding more motors proportionally increases the system velocity. As $N$ grows much larger than $\Dm/\Dc$, however, the steady-state velocity asymptotically approaches maximum velocity $v_\mathrm{max} =\beta \Dm f_\mathrm{chem}$ (the mean velocity of an uncoupled motor) as $\langle v\rangle \approx v_\textrm{max}\left[1 - \Dm/(N\Dc)\right]$.
Thus no matter how many motors are coupled to the cargo, the mean velocity of the aggregate motor-cargo system never exceeds that of an unladen motor. This mean velocity (as well as the maximum velocity) scales linearly with the chemical driving force $\force$. Figure~\ref{fig:properties}a shows normalized velocity $\langle v\rangle/v_{\rm max}$ as a function of $N$.

While $\langle v\rangle$ gives the average motion, our model (like all transport at the cellular level) is inherently stochastic. As such, the average velocity is not sufficient to fully describe system behavior, even at steady state. The effective cargo diffusivity $D_\mathrm{eff}$ quantifies the rate at which the variance of forward progress grows at steady state:
\begin{subequations}\label{eq:D_eff}
\begin{align}
D_\mathrm{eff} & \equiv \lim_{t\to\infty} \frac{\langle \xc^2\rangle -\langle \xc\rangle^2}{2t}\\
& = \left(\frac{1}{\Dc} + \frac{N}{\Dm}\right)^{-1}.
\end{align}
\end{subequations}
This expression can be understood by noting that the effective friction coefficient for the system, $\zeta_\mathrm{eff}=1/\beta D_\mathrm{eff}$, is simply the sum of the friction coefficients for the motors and cargo. This interpretation is consistent with a recent theoretical study of collective transport of a diffusive cargo using discrete motor dynamics, which similarly found that the contribution from the motors to the effective friction coefficient of the system scales linearly with the number $N$ of motors for the small range explored~\cite{nakul2021frictional}. Previous work using a simpler phenomenological model also suggested that the effective friction coefficient of a collection of motors should be proportional to $N$~\cite{leibler1993porters}. 

Figure~\ref{fig:properties}a shows $D_\mathrm{eff}$ as a function of $N$. While the velocity increases with the number of motors, the effective diffusivity decreases, indicating that a larger number of motors tightens the distribution of cargo-transport distances. Writing the mean velocity as $\langle v\rangle = \beta D_\mathrm{eff} \left(N\force\right)$ reveals that the effective dynamics of the system are simply those of a single diffusive particle (with diffusivity $D_\mathrm{eff}$) under a constant driving force $N\force$.

The system stochasticity can alternatively be quantified by the coefficient of variation $\CV$~\cite{brown1998coefficient} or the Fano factor $\phi$~\cite{brown2019pulling}. The coefficient of variation (CV) of cargo position is
\begin{subequations}\label{eq:CoV}
\begin{align}
\CV & \equiv \frac{\sqrt{\langle x_\mathrm{c}^2\rangle - \langle x_\mathrm{c}\rangle^2}}{\langle x_\mathrm{c}\rangle} \\
& = \sqrt{2\left(\frac{1}{\Dc} + \frac{N}{\Dm}\right)}\, \frac{\Dc}{N\Dm}\frac{1}{\beta f_\mathrm{chem}}\, t^{-1/2}.
\end{align}
\end{subequations}
For small $N\ll \Dm/\Dc$ the coefficient of variation scales as $N^{-1}$, while for large $N\gg\Dm/\Dc$, $\CV\propto N^{-1/2}$. Thus this measure of the variation in forward progress can be made arbitrarily small with sufficiently large $N\Dc/\Dm$, but with diminishing returns for larger $N$.

The steady-state Fano factor is
\begin{equation}\label{eq:fanofactor}
\fano \equiv \frac{
\langle x_\mathrm{c}^2\rangle - \langle x_\mathrm{c}\rangle^2}{\langle \xc\rangle} = \frac{2}{N\beta\force}.
\end{equation}
Similarly to $D_\mathrm{eff}$, $\fano$ decreases (and hence the precision increases) with the number of motors, scaling as $\fano\propto N^{-1}$. Here adding motors decreases the variance of forward progress while increasing the velocity, leading to a Fano factor that decreases with $N$. 
For $N=1$ motor, Eq.~\eqref{eq:fanofactor} recovers the Fano factor previously calculated in the limit of low cargo diffusivity~\cite{brown2019pulling} for a single motor. 

The addition of motors can be thought of as having an ``averaging" effect on the dynamics. The precision (as quantified by $\CV$ or $\fano$) also increases with the chemical driving force $\force$ on each motor. Figure~\ref{fig:stochastic_metrics} in Appendix~\ref{stochasticity_app} shows the three stochasticity metrics.

\subsubsection{Thermodynamic properties}
We also exactly calculate steady-state ensemble averages of all energy flows into and out of each subsystem. The mean chemical power~\eqref{motor_power_eq} to each motor is
\begin{subequations}\label{eqn:P_to_M}
\begin{align}
\beta \PMi & = N\Dc \left(1 + N\frac{\Dc}{\Dm}\right)^{-1} \left(\beta\force\right)^2 \\
& = \beta \force \langle v\rangle.   
\end{align}
\end{subequations}
We multiply the energy flows by the inverse temperature $\beta$ so that quantities have units of $\mathrm{s}^{-1}$. In keeping with our assumption of tight mechanochemical coupling, the power consumption is simply the chemical driving force multiplied by the motor velocity. Since the motors are identical, the total chemical power consumption is simply $P_\text{chem} = N \PMi$. Likewise, the average rate of work performed on the cargo by each motor is
\begin{subequations}\label{eqn:P_M_to_C}
\begin{align}
\beta \PMC & = N \Dc \left(1 + N\frac{\Dc}{\Dm}\right)^{-2}\left(\beta\force\right)^2 \\
& = \frac{\langle v\rangle^2}{N\Dc}. 
\end{align}
\end{subequations}
The total power $\sum_i \PMC = \langle v\rangle^2/\Dc$ transferred from the motors to the cargo increases monotonically with $N$ and reaches a finite maximum value. $\PMC$ for a single motor is non-monotonic in $N$, as shown in Fig.~\ref{fig:properties}b. In particular, $\PMC\propto N$ for $N\ll \Dm/\Dc$ and $\propto N^{-1}$ for $N\gg \Dm/\Dc$. This is because for large $N$, when the cargo reaches maximum velocity and thus a constant rate of heat dissipation, the sum $\sum_i \PMC = -\QC$ must reach a constant value as well. Dividing this nearly constant total power among an increasing number of motors means that $\PMC$ decreases. Thus the power flow from each individual motor to the cargo is maximized at an intermediate $N^*=\Dm/\Dc$. 

The average heat flow into each motor is
\begin{subequations}\label{eqn:Q_to_M}
\begin{align}
\beta\QMi & = -N^2\frac{\Dc^2}{\Dm} \left(1 + N\frac{\Dc}{\Dm}\right)^{-2}\left(\beta\force\right)^2\\
& = -\frac{\langle v\rangle^2}{\Dm},
\end{align}
\end{subequations}
and the heat flow into the cargo is
\begin{subequations}\label{eqn:Q_to_C}
\begin{align}
\beta \QC & = -N^2\Dc \left(1 + N\frac{\Dc}{\Dm}\right)^{-2}\left(\beta\force\right)^2\\
& = -\frac{\langle v\rangle^2}{\Dc}. 
\end{align}
\end{subequations}
The subsystem-specific heat flows are, on average, given by the friction coefficient (e.g. $\zeta_\mathrm{c}=1/\beta \Dc$ for the cargo) multiplied by the mean velocity squared, the result we would expect for the frictional energy dissipation of an overdamped particle moving at constant velocity $\langle v\rangle$. The sum of the heat flows over all $N+1$ subsystems represents the total energy dissipation of the system at steady state. As indicated by Eq.~\eqref{second_law_heat}, all of the chemical energy consumed by the motors is either dissipated directly by the motors as heat (due to loose coupling to the cargo) or transduced into mechanical work on the cargo which is then dissipated by the cargo as heat.

Figure~\ref{fig:properties}b shows how these four steady-state energy flows depend on the number $N$ of motors, manifesting two regimes. For $N\ll \Dm/\Dc$, the heats $-\QMi$ and $-\QC$ scale as $N^2$, while the chemical power $\PMi$ to each motor and the power $\PMC$ from each motor to the cargo scale linearly with $N$. For $N\gg \Dm/\Dc$, the chemical power to each motor as well as the two heats asymptotically approach constants. For sufficiently large $N$, each motor's heat roughly equals its chemical power consumption, as the power per motor transferred through the coupling decays as $N^{-1}$.

All energy flows display the same quadratic dependence on the chemical driving force. This is reminiscent of linear irreversible thermodynamics, where rates of entropy production (and thus heat dissipation) are quadratic in the thermodynamic driving forces~\cite{de1969non}. This is true for our system on average due to the linearity of Eqs.~\eqref{multiplemotorlangevin}a-b, even with the inherent stochasticity. 

The steady-state energy flows~\eqref{eqn:P_to_M}-\eqref{eqn:Q_to_C} are all independent of the coupling strength $\kappa$, despite the steady-state distributions for the separation distances $\ri$ depending strongly on $\kappa$. To understand this surprising result, consider the average force an individual motor pulls against, $\kappa \langle \ri\rangle$, for separation distance $\ri\equiv \xmi-\xc$. Equation~\eqref{eq:mean_r} shows that $\langle \ri\rangle\propto 1/\kappa$, so the mean force on the motor is independent of the coupling strength. Since the motor velocity is also independent of the coupling strength, the mean power output of each motor (roughly the mean velocity multiplied by the mean opposing force) is independent as well. The power consumption $P_{\to \mathrm{M}_i}$ is likewise independent of $\kappa$ for the same reason. From the first law \eqref{first_law} the system only has three independent energy flows, so the other energy flows must thus also be independent of $\kappa$.
 
We consider several different metrics of steady-state energetic efficiency. Thermodynamic efficiency is the ratio of output power to input power. Since the system does not perform any thermodynamic work as output, and the input power $P_\text{chem}$ is always positive, the full system's thermodynamic efficiency is zero; however, the thermodynamic efficiency $\etaM \equiv \PMC/\PMi$ of each motor subsystem is positive.

The Stokes efficiency $\etaS \equiv \zeta_{\rm c} \langle v\rangle^2/\Pchem$ evaluates systems whose only functional output is directed motion against viscous friction~\cite{wang2002stokes}. We consider only the frictional drag on the cargo (with drag coefficient $\zeta_{\rm c}$), assuming that directed cargo motion is ultimately the main purpose of the system. This system has equal Stokes efficiency and motor efficiency:
\begin{equation}\label{effic_eq} 
\etaM = \etaS = \left(1 + N\frac{\Dc}{\Dm}\right)^{-1}.
\end{equation}
Figure~\ref{fig:properties}a shows efficiency as a function of $N$: for small $N\Dc/\Dm$, the efficiency is $\approx 1- N\Dc/\Dm$, while for $N\Dc/\Dm\gg1$ the efficiency scales as $N^{-1}$. For a given diffusivity ratio $\Dc/\Dm$, the efficiency for any $N$ is upper bounded by $\eta_\text{max} = \left(1 +\Dc/\Dm\right)^{-1}$ since $N$ is lower bounded by unity. Thus, for example, a system with $\Dc=\Dm$ can achieve at most $50\%$ efficiency. This efficiency can be re-written in terms of the system's effective diffusivity as
\begin{equation}
\etaM = \etaS = \frac{D_\mathrm{eff}}{\Dc},
\end{equation}
which exactly saturates an upper bound proven for the Stokes efficiency of transport by a single motor~\cite{pietzonka2016universal}.

Table~\ref{tab:table1} summarizes the scaling with $N$ of our key performance metrics in the two limiting regimes.

\begin{table}[h] 
\caption{\label{tab:table1}
Performance metrics' asymptotic scaling with $N$.}
\vspace{1mm}
\bgroup
\def\arraystretch{1.6}
\begin{ruledtabular}
\begin{tabular}{c|cc}
\textrm{Metric}&
$N\ll \Dm/\Dc$&
$N\gg \Dm/\Dc$\\
\colrule
$\langle v\rangle$ & $\propto N$ & $\approx v_\textrm{max}\left[1 - \Dm/(N\Dc)\right]$\\ 
$\CV$ & $\propto N^{-1}$ & $\propto N^{-1/2}$\\
$P_\textrm{chem}$ & $\propto N^2$ & $\propto N$\\
$\eta_{\mathrm{S}/\mathrm{M}}$ & $\approx 1- N\Dc/\Dm$ & $\propto N^{-1}$\\
\end{tabular}
\end{ruledtabular}
\egroup
\end{table}

\subsection{Performance trade-offs}\label{sec:tradeoffs}
The previous section outlined the separate $N$-dependence of different performance metrics; however, these quantities are not independent, instead posing trade-offs parameterized by $N$. We examine the trade-offs between all pairs of dynamical and thermodynamic properties from Table~\ref{tab:table1}, and find that several pairs of desirable properties cannot be simultaneously attained.

\begin{figure*}[t]
\includegraphics[width=\textwidth]{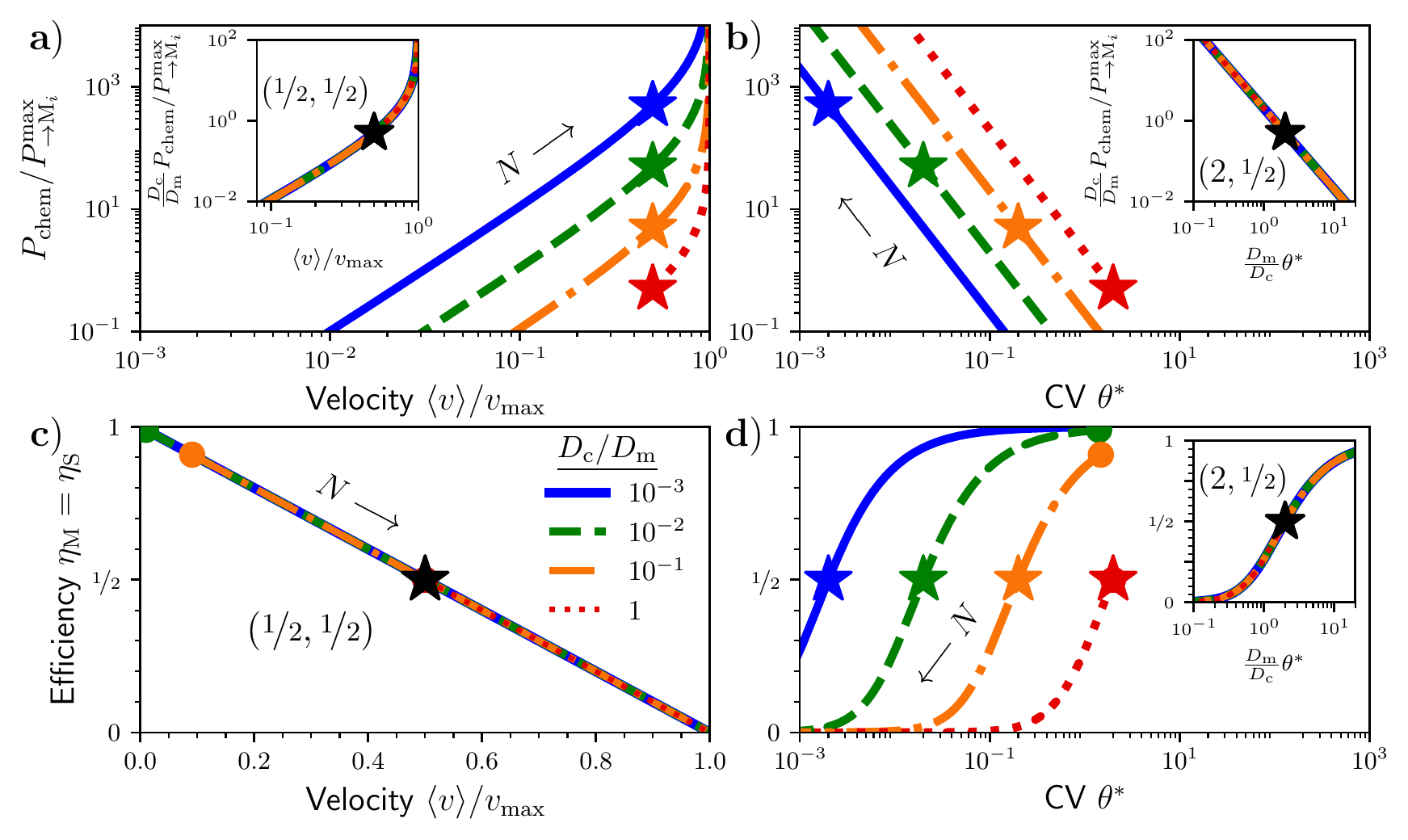} 
\caption{\label{fig:tradeoffs} Trade-offs between \textbf{a)} scaled chemical power consumption $\Pchem/\PMi^\mathrm{max}$ and scaled mean velocity $\langle v\rangle/v_\mathrm{max}$, \textbf{b)} $\Pchem/\PMi^\mathrm{max}$ and scaled coefficient of variation $\CV^* = \CV v_\mathrm{max}\sqrt{t/D_\mathrm{c}}$, \textbf{c)} efficiency $\etaS=\etaM$ and $\langle v\rangle/v_\mathrm{max}$, and \textbf{d)} $\etaS=\etaM$ and $\CV^*$, as the number $N$ of motors is varied, for different diffusivity ratios $\Dc/\Dm$. Stars: $N=\Dm/\Dc$. Circles: maximum efficiencies for respective diffusivity ratios, $\eta_\mathrm{max}=\left(1 + \Dc/\Dm\right)^{-1}$, realized for $N=1$. Insets in \textbf{a)}, \textbf{b)}, and \textbf{d)} show that scaling the power consumption and coefficient of variation by factors of $\Dc/\Dm$ collapses the curves for different diffusivity ratios onto single master curves. Numbers in parentheses indicate coordinates of black stars (where $N\Dc/\Dm=1$). All curves terminate at points where $N=1$, so that we always show real numbers of motors.}
\end{figure*}

The mean transport velocity and the total chemical power consumption of the motors are related by
\begin{equation}
\frac{\Pchem}{P_{\to \mathrm{M}_i}^\mathrm{max}} = \frac{\Dm}{\Dc}\frac{\left(\langle v\rangle/v_\mathrm{max}\right)^2}{1 - \langle v\rangle/v_\mathrm{max}}.
\end{equation}
Figure~\ref{fig:tradeoffs}a illustrates this trade-off as $N$ is varied, for several different diffusivity ratios. For $N\ll \Dm/\Dc$, the total chemical input power scales as the square of the average velocity $\langle v\rangle$. At $N=\Dm/\Dc$ the velocity is half its maximum; beyond this velocity the required chemical power skyrockets, scaling as $\Pchem\propto \left(v_\mathrm{max}-\langle v\rangle\right)^{-1}$ for $N\gg \Dm/\Dc$.

The total power consumption and coefficient of variation~\eqref{eq:CoV} are inversely related, 
\begin{equation}
\Pchem = (2/t) \CV^{-2},
\end{equation}
for all $N$ and $\Dc/\Dm$. Figure~\ref{fig:tradeoffs}b illustrates this trade-off, which features scaling behavior consistent across the regimes of large and small $N$. Arbitrarily high precision (low $\CV$) can be achieved in this collective-transport system, but at the cost of power consumption that increases without bound.

Comparing \eqref{velocityeq} and \eqref{effic_eq}, the efficiency and scaled velocity obey a simple relation:
\begin{equation}
\eta_{\mathrm{S}/\mathrm{M}} + \frac{\langle v\rangle}{v_\text{max}} = 1,
\end{equation}
where $\eta_{\mathrm{S}/\mathrm{M}}$ can be either the Stokes or motor efficiency, since they are equal for this system. Fig~\ref{fig:tradeoffs}c shows this trade-off. Our collective-transport system cannot simultaneously achieve high efficiency and near-maximum velocity. Further, depending on the diffusivity ratio only certain efficiencies are achievable: those with $\eta_{\mathrm{S}/\mathrm{M}}$ $\leq\eta_\mathrm{max}=\left(1 + \Dc/\Dm\right)^{-1}$. Notably, $50\%$ efficiency and half-maximal velocity can always be achieved at $N = \Dm/\Dc$.

Finally we consider the trade-off between efficiency and precision (quantified by coefficient of variation $\CV$),
\begin{equation}
\CV \, v_\mathrm{max} \sqrt{\frac{t}{2\Dc}} = \frac{\sqrt{\eta_{\mathrm{S}/\mathrm{M}}}}{1-\eta_{\mathrm{S}/\mathrm{M}}}.
\end{equation}
Figure~\ref{fig:tradeoffs}d shows that high efficiency and high precision (low CV) are not simultaneously achievable. This suggests that to maximize efficiency systems must exploit thermal fluctuations, leading to a decrease in precision. Note that the transition from near-zero to near-unit efficiency occurs over a small range of CVs around $N=\Dm/\Dc$.

Insets in Figs~\ref{fig:tradeoffs}a, b, and d show that scaling the power consumption and coefficient of variation by factors of the diffusivity ratio $\Dc/\Dm$ collapses the separate curves for distinct $\Dc/\Dm$ onto single master curves. Thus the qualitative nature of the trade-offs described here are independent of the relative diffusivities of the motors and cargo.

These performance trade-offs suggest that collective-transport systems where different performance metrics are prioritized should have different numbers of motors if $N$ can be adjusted to tune performance. For systems in which maximum velocity and high precision are preferred, optimization would drive systems towards the $N\gg \Dm/\Dc$ regime. This would however come with the cost of high power consumption and decreased efficiency. If instead highly efficient directed transport on a small power budget is favored, then optimal systems would have $N\ll \Dm/\Dc$ at the cost of slow and imprecise transport. 

\subsection{Generalization}
Many of our results extend to more general models of motor dynamics. As an example, we relax the assumption that the chemical driving force is much larger than the scale of the motor's energy landscape, and add a periodic sinusoidal potential-energy landscape to each motor (see Appendix \ref{simulation_data} for details). This continuous model, inspired by similar models of other molecular machines~\cite{zimmermann2015effective,lathouwers2020nonequilibrium}, produces motor dynamics qualitatively similar to commonly used discrete-step models~\cite{bhat2016transport,mckinley2012asymptotic}.

Figure~\ref{fig:simulationscalingfig} shows 
for this more complex model
the scaling with $N$ of the mean velocity $\langle v\rangle$, chemical power consumption $P_\mathrm{chem}$, and Stokes efficiency $\etaS$, for a variety of barrier heights. 
(Calculating the coefficient of variation for large $N$ is computationally intractable.) 
The scaling laws in the limiting regimes of large and small $N$ (outlined in Table 1) still accurately reflect the limiting scaling behavior for this generalization. 
As a direct result of the scaling laws generalizing, the performance trade-offs [Figs.~\ref{fig:tradeoffs}a and c] also apply more generally, at least qualitatively: even under more general motor dynamics, desirable pairs of properties such as high velocity and high efficiency or high velocity and low power consumption remain mutually exclusive.

\section{Discussion}\label{discussion}
In this article we introduced a simple model for collective intracellular transport by an arbitrary number of transport motors. This model is stochastic, thermodynamically consistent, and can be solved analytically for arbitrary motor number $N$. Using this model we derived analytic expressions for several steady-state properties, including dynamic properties such as mean velocity, effective diffusivity, and precision, as well as thermodynamic quantities such as power, heat, and efficiency. We found qualitatively different $N$-dependence for these quantities in the two opposing regimes of high and low $N$ (compared to the motor-cargo diffusivity ratio $\Dm/\Dc$), summarized by simple scaling laws (Table~\ref{tab:table1}). Our model should best reflect the physics of motors whose energy-landscape features are smaller in scale than the magnitude of their chemical driving, however our numerical explorations in Appendix~\ref{simulation_data} suggest that many of our results generalize well beyond this regime.

We also examined trade-offs between several performance metrics that are expected to be generally important for real transport systems: many pairs of desirable properties, for example high velocity and high efficiency, are mutually exclusive in these systems. The trade-off between efficiency and velocity (Fig.~\ref{fig:tradeoffs}c) has also been explored theoretically in other types of molecular machines~\cite{wagoner2019opposing,wagoner2021evolution}. The incompatibility of high velocity and high efficiency seems to be a general feature of these types of systems, and has also been seen experimentally for myosin motors~\cite{purcell2011nucleotide}. These findings are reminiscent of the thermodynamic uncertainty relation~\cite{horowitz2020thermodynamic}, which lower bounds the product of uncertainty and entropy production. At steady state our system saturates this bound, which may be a universal feature of linear systems with only one driving force \cite{barato2015thermodynamic} or systems described by Gaussian probability distributions \cite{saryal2019thermodynamic}. This suggests that the performance trade-offs (especially Fig.~\ref{fig:tradeoffs}b) may in fact be Pareto frontiers~\cite{shoval2012evolutionary} for more general collective-transport systems.

Few experiments have measured exact numbers of motors in collective transport systems; nonetheless, motor numbers have indeed been measured, for example by identifying discrete peaks in a distinctly multimodal velocity distribution. Using this technique to study transport of beads by kinesin motors, \cite{shtridelman2008force,shtridelman2009vivo} found a mean velocity roughly proportional to the number of motors, consistent with our predictions for the small-$N$ regime; the bead diameter and solution viscosity indicate that these experiments had $N<\Dc/\Dm$. Similar experimental investigations have also found velocity to be a concave function of $N$, qualitatively consistent with our predictions~\cite{gagliano2010kinesin}. Other experiments have varied the concentration of motor proteins in solution, a rough proxy for the number of motors per cargo. Kinesin motors attached to a substrate while transporting long microtubules~\cite{kaneko2019transport,howard1989movement} produce velocity that is a concave function of motor concentration, as we predict; similar results have been found for myosin motors pulling actin filaments~\cite{rastogi2016maximum}.

Using models similar to ours, but with discrete steps rather than continuous motor dynamics, \cite{korn2009stochastic} and \cite{bhat2016transport} found a similar monotonic and concave functional dependence of the mean velocity $\langle v\rangle$ on motor number $N$, although their findings are restricted to relatively small $N$. Likewise, \cite{bhat2016transport} also found $D_\mathrm{eff}\propto1/N$ for the greatest $N$ they investigated. Another recent study~\cite{bhat2017stall} found that the total load capacity, or effective stall force, is $N$ times the stall force for a single motor. We can easily incorporate into our model a constant external force $f_\mathrm{ext}$ pulling the cargo in the opposite direction of the chemical force driving the motors. To do this we simply add a term $-\beta \Dc f_\mathrm{ext}$ to the left side of \eqref{eq:cargoMotion}, preserving analytic solubility of the dynamical equations. The mean velocity in this case is
\begin{equation}
\langle v\rangle_{f_\mathrm{ext}} = \langle v\rangle_0 \left(1 - \frac{f_\mathrm{ext}}{N f_\mathrm{chem}}\right),
\end{equation}
where $\langle v\rangle_0$ is the mean velocity for $f_\mathrm{ext}=0$ given by Eq.~\eqref{velocityeq}. From this we identify the stall force $f_\mathrm{s} = Nf_\mathrm{chem}$ which scales linearly with the number of motors, independent of the diffusivity ratio.

Throughout this article we treated the number $N$ of active motors per cargo as fixed, despite the fact that in real transport systems motor proteins are constantly binding and unbinding to both the cargo and the substrate. Our estimate $\tau_\mathrm{relax}<0.02$\,s of our system's relaxation time (Sec.~\ref{Solution_timescale}) is much shorter than estimates of $0.2-1$\,s for the motor binding and unbinding timescales of kinesin on microtubules~\cite{klumpp2005cooperative}. Due to this timescale separation, we can treat the system as always in dynamical steady state at fixed $N$ even when motors bind and unbind over longer timescales. Thus our steady-state results should still hold for temporally varying $N$. The convexity of these properties with respect to $N$ determines the sign of the error resulting from computing steady-state quantities at a single average motor number $\langle N\rangle$ instead of considering a full distribution $P(N)$. For example, mean velocity is a concave function of $N$, so when $N$ varies $\left\langle v \left(\langle N\rangle\right)\right\rangle$ overestimates the mean velocity $\langle v(N) \rangle$. By contrast, the total power consumption is convex, so $\langle \Pchem(N)\rangle \geq \Pchem\left(\langle N\rangle\right)$. In Appendix~\ref{fluctuating_N} we use a simple stochastic dynamical model of motor binding/unbinding to estimate that the error due to treating $N$ as constant is small, especially when $N$ is far from $\Dm/\Dc$.

Our model also ignores other possibly relevant effects such as motor-motor interactions and discretization of motor steps. Depending on the time- and lengthscales of resolution, motor proteins like kinesin can be thought of as taking discrete steps~\cite{svoboda1993direct}, in contrast to the continuous motion we assumed here. Nevertheless, treating the motor dynamics as continuous should be valid as long as the relaxation timescale for the motors is sufficiently separated from that of the cargo (either much larger or much smaller) so that the steady-state separation distance distribution~\eqref{eq:r_dist} converges to the resulting equilibrium distributions. This assumption is valid for both $N\gg \Dm/\Dc$ and $N\ll \Dm/\Dc$, but may break down for $N\approx \Dm/\Dc$, where system behavior may be more sensitive to the exact motor dynamics.

Interactions between motors may become relevant under certain conditions. Computational studies indicate the possibility of long-range cooperative interactions between kinesin motors through microtubules~\cite{wijeratne2020motor} as well as crowding effects such as traffic jams when large numbers of motors are present~\cite{klumpp2005movements}. While we have not incorporated these effects into our model at present, generalizing our results using model-agnostic considerations from the theory of stochastic thermodynamics is a promising future direction.

Our model does break down under certain limiting conditions. Our analysis yields a mean velocity proportional to the chemical driving force acting on each motor. For kinesin motors hydrolyzing ATP, this is consistent with experimental findings at lower ATP concentrations, but not at high ATP concentrations where motor velocity saturates, with a cross-over at ATP concentrations on the order of 0.1mM~\cite{schnitzer1997kinesin} (depending on the load the motor is pulling against). This discrepancy arises because we do not explicitly model the motor's mechanochemical cycle, implicitly treating the binding of ATP as the rate-limiting step. Reaction steps not dependent on the free-energy consumption should become rate-limiting at high ATP concentrations~\cite{carter2005mechanics} leading to a saturating velocity. Our main results (the $N$-dependence  and trade-offs of key performance metrics) are all independent of the magnitude of $\force$. Nonetheless, we expect our model to best capture the dynamics of real systems at lower ATP concentrations, in the linear-velocity regime. For systems with very large chemical driving forces, we expect the true maximum velocity of a single uncoupled motor to saturate, unlike our predicted $v_\mathrm{max}=\beta\Dm\force$.

The performance trade-offs we derived in Sec.~\ref{sec:tradeoffs} point to insights about optimization in collective-transport systems, as adjusting the number of motors per cargo can tune performance. This could be achieved, for example, by manipulating the motor concentration~\cite{ndlec1997self}, adjusting the number of possible binding sites on the cargo~\cite{kumar1995kinectin}, or using extra structural assemblies such as DNA scaffolds~\cite{furuta2013measuring}. Depending on the regime the system inhabits, as determined by the dimensionless quantity $N\Dc/\Dm$, systems can either achieve fast and precise but energetically costly transport ($N\gg \Dm/\Dc$), or efficient but slow and imprecise transport ($N\ll \Dm/\Dc$). Ultimately, real systems have likely evolved to optimize complex combinations of the performance metrics we have considered here and others we have not, however estimating $N\Dc/\Dm$ for \textit{in vivo} systems may provide insight into which performance metrics are most important in specific systems.

\begin{acknowledgments}
The authors thank Emma Lathouwers and Jannik Ehrich (SFU Physics) for useful discussions, and Eric Jones and Nancy Forde (SFU Physics) for feedback on the manuscript. We thank the two anonymous reviewers whose comments and suggestions helped improve and clarify this article. This work was supported by a Natural Sciences and Engineering Research Council of Canada (NSERC) CGS Masters fellowship (M.P.L.), a BC Graduate Scholarship (M.P.L.), an NSERC Discovery Grant and Discovery Accelerator Supplement (D.A.S.), and a Tier-II Canada Research Chair (D.A.S.).
\end{acknowledgments}

\bibliography{main}

\appendix
\counterwithin{figure}{section}

\section{Analytic expressions for means and covariances}\label{ana_mu_cov}

As detailed in Sec.~\ref{Solution_timescale}, the coupled Langevin equations (Eq.~\eqref{multiplemotorlangevin}a and b) that describe the dynamics of our system are analytically solvable, and the solution for the probability distribution of the respective positions $\xc$ and $\{\xmi\}_{i=1}^N$ of the cargo and motors is a multivariate Gaussian. Starting from initial conditions $\xc = \xmi = 0$ at $t=0$, at steady state ($t\gg\tau_\mathrm{relax}$) the mean cargo position and motor positions are
\begin{subequations}\label{eq:means}
\begin{align} 
\langle \xc\rangle & = \langle v\rangle t - \frac{N\force}{\kappa}\frac{D_\mathrm{eff}^2}{\Dm\Dc},\\
\langle \xmi\rangle & = \langle v\rangle t + \frac{\force}{\kappa}\left(\frac{D_\mathrm{eff}}{\Dc}\right)^2,
\end{align}
\end{subequations}
and the covariances are
\begin{subequations}\label{eq:covariances}
\begin{align}
\mathrm{Cov}(\xc,\xc) & = 2D_\mathrm{eff}t + \frac{N}{\beta\kappa}\left(\frac{D_\mathrm{eff}}{\Dm}\right)^2,\\
\mathrm{Cov}(\xc,\xmi) & = 2D_\mathrm{eff}t - \frac{1}{\beta\kappa}\frac{D_\mathrm{eff}^2}{\Dm\Dc},\\
\mathrm{Cov}(\xmi,\xmj) & = 2D_\mathrm{eff}t \\
& \quad + \frac{1}{\beta\kappa} \left[\delta_{ij} - \frac{D_\mathrm{eff}^2}{\Dm}\left( \frac{N}{\Dm} + \frac{2}{\Dc}\right)\right],\nonumber
\end{align}
\end{subequations}
for Kronecker delta function $\delta_{ij}$. Different initial conditions would produce different time-independent constant terms in \eqref{eq:means} and \eqref{eq:covariances}; however for large times (in the steady-state limit) the constant terms are negligible compared to the terms linear in $t$. Regardless of initial conditions, the difference between the constant terms in \eqref{eq:means}a and b will always be the mean value of the separation distance, $\langle \ri\rangle$.

\section{Parameter estimates}\label{params}
In this Appendix we discuss relevant experimental measurements of the parameters in our model, namely the linker spring constant $\kappa$, the chemical driving force $\force$, and the respective motor and cargo diffusivities $\Dm$ and $\Dc$.

Experimentally, kinesin linkers are well-approximated as Hookean springs with a zero rest length and a spring constant $\sim$$0.5\,\mathrm{pN/nm}$~\cite{kawaguchi2003equilibrium}. Similar behavior has been observed for the linkers of myosin V motors, which have an estimated spring constant of $0.2-0.4\,\mathrm{pN/nm}$~\cite{veigel2005load}.

We estimate the chemical driving force $\force$ in two ways. By noting the equivalence in Eq.~\eqref{force_velocity} of $\force$ with the single-motor stall force, we use experimental estimates of single-motor stall forces to estimate the chemical driving force. Kinesin motors have stall forces on the order of $5-8\,\mathrm{pN}$~\cite{visscher1999single}, while myosin motors stall at forces as high as $15\,\mathrm{pN}$~\cite{debold2005slip}. (Note that the stall force monotonically increases with cellular ATP concentration.) 

Likewise, due to tight coupling between chemical energy consumption and mechanical motion~\cite{schnitzer1997kinesin,visscher1999single}, $\force$ can also be thought of as a free-energy dissipation per unit distance. Kinesin, for example, hydrolyzes one molecule of ATP (a reaction with a free-energy change $\Delta \mu_\mathrm{ATP}\approx 15-30 \,k_\mathrm{B}T$~\cite{milo2015cell}) for every forward step ($d\approx 8\,\mathrm{nm}$). At 298 K, $1\,k_\mathrm{B}T = 4.114\,\mathrm{pN}\,\mathrm{nm}$, resulting in a chemical driving force $\force = \Delta \mu_\mathrm{ATP}/d \approx 8-15\,\mathrm{pN}$, in line with our previous estimate.

We estimate the motor diffusivity using $v_\mathrm{max} = \beta f_\mathrm{s}\Dm$. For kinesin-1 the maximum velocity is $v_\mathrm{max}\approx 1-2\,\mu\mathrm{m/s}$ and the stall force is $f_\mathrm{s}\approx 6-8\,\mathrm{pN}$~\cite{fallesen2011force}, while for myosin V, $v_\mathrm{max}\approx 8\,\mathrm{nm/s}$~\cite{purcell2011nucleotide} and $f_\mathrm{s} \approx 10-15\,\mathrm{pN}$~\cite{rastogi2016maximum}. This suggests that in both cases $\Dm = \mathcal{O}(10^{-3})\,\mu\mathrm{m}^2/\mathrm{s}$. Alternatively, using experimental estimates of rate constants for forward and reverse steps yields an estimate for kinesin-1 motor diffusivity of $\Dm \approx 4\times 10^{-3}\,\mu\mathrm{m}^2/\mathrm{s}$~\cite{vu2016discrete}. 

Cargo diffusivity can vary by orders of magnitude depending on the type of molecular cargo. As one example, diffusivity of vesicles (with radii on the order of $300\,\mathrm{nm}$) in neurons is estimated to be of order $10^{-3}\,\mu\mathrm{m}^2/\mathrm{s}$~\cite{ahmed2014active}. Other measurements of vesicles and vesicle-sized beads in cytoplasm have found diffusivities on the order of $10^{-4}-10^{-2}\,\mu\mathrm{m}^2/\mathrm{s}$~\cite{luby1999cytoarchitecture}. Larger cargo such as organelles, for example mitochondria which have diameters as large as $2\mu\rm m$, will have even smaller diffusivities. Thus for intracellular transport of vesicles and organelles we expect $\Dc/\Dm\in[10^{-3},1]$.

We use these parameter ranges to estimate the relaxation time $\tau_\mathrm{relax} = \left[ \beta\kappa (\Dm + N\Dc)\right]^{-1}$ in Sec.~\ref{Solution_timescale}. Based on the ranges and estimates above, we get a maximum value of about $0.02$\,s (taking $\kappa=0.2\,\mathrm{pN/nm}$, $T=298\,\mathrm{K}$, $\Dm=10^{-3}\,\mu\mathrm{m}^2/\mathrm{s}$, $\Dc=10^{-4}\,\mu\mathrm{m}^2/\mathrm{s}$, and $N=1$). Using more typical values of these parameters (for example $\kappa=0.5\,\mathrm{pN/nm}$, $\Dm=4\times10^{-3}\,\mu\mathrm{m}^2/\mathrm{s}$, $\Dc=10^{-3}\,\mu\mathrm{m}^2/\mathrm{s}$, and $N=10$) gives a much smaller estimate $\tau_\mathrm{relax}\approx 5\times 10^{-4}$\,s.

\section{Stochasticity metrics}\label{stochasticity_app}
\begin{figure}[h]
\includegraphics[width=\columnwidth]{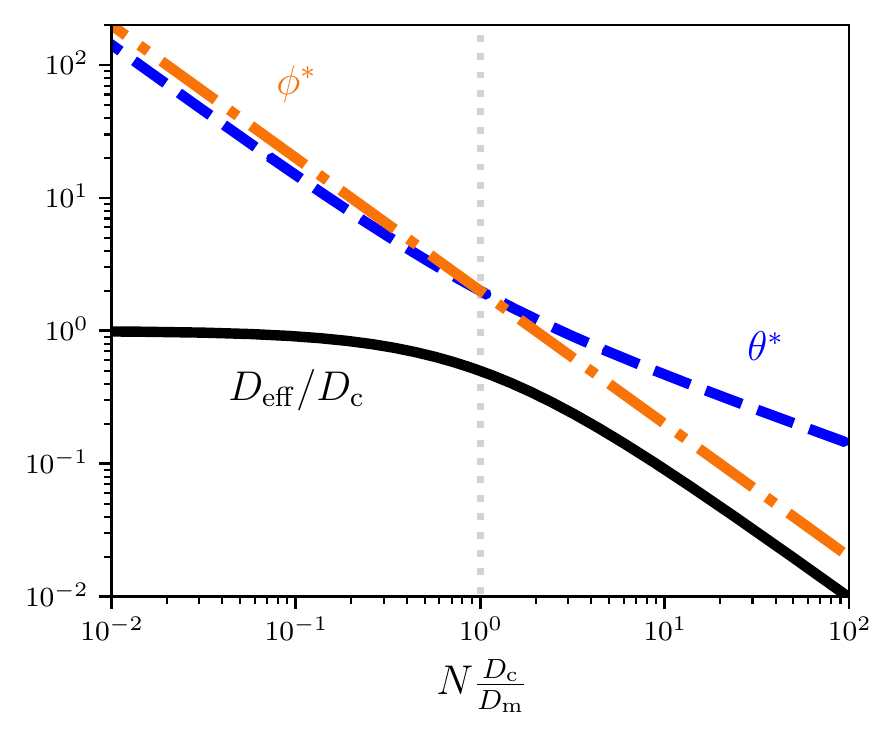}
\caption{\label{fig:stochastic_metrics} $N$-dependence of the performance metrics used to quantify stochasticity: the effective diffusivity $D_\mathrm{eff}/D_\mathrm{c}$, the coefficient of variation $\CV^* = \CV \beta f_\mathrm{chem}\sqrt{t/D_\mathrm{c}}$, and the Fano factor $\fano^* = \beta \force (\Dm/\Dc) \fano$; all in dimensionless units.}
\end{figure}

Figure \ref{fig:stochastic_metrics} shows the $N$-dependence of the three stochasticity metrics we introduced in Sec.~\ref{sec:dynamical}: the effective diffusivity $D_\mathrm{eff}$~(Eq.~\eqref{eq:D_eff}), the coefficient of variation $\CV$~(Eq.~\eqref{eq:CoV}), and the Fano factor $\phi$~(Eq.~\eqref{eq:fanofactor}).

\section{Simulations for motors with significant energy barriers}\label{simulation_data}
We add to each motor a periodic potential-energy landscape of the form $V(\xmi) = E^\ddagger \cos \left(\xmi/\ell\right)$, where $2E^\ddagger$ is the height of the energy barriers between successive meta-stable states (local energy minima), and $2\pi \ell$ is the period. Each motor's dynamics satisfy
\begin{equation}\label{eq:modifiedmotorMotion}
\dotxmi = \beta \Dm\left[\force - \kappa (\xmi - \xc) - f_\mathrm{max}\sin(\xmi/\ell)\right] + \eta_{i},
\end{equation}
where $f_\mathrm{max} = E^\ddagger/\ell$ is the maximum conservative force arising from the periodic potential. (The cargo motion still obeys Eq.~\eqref{eq:cargoMotion}.) While these new equations of motion cannot be solved analytically, we can numerically simulate the dynamics of this system, by integrating the $N+1$ Langevin equations for a given value of $N$. Obtaining full time-dependent probability distributions through simulation is computationally intractable for large $N$, so we compute only properties that depend solely on the average system dynamics.

\begin{figure}[!ht]
\includegraphics[width=\columnwidth]{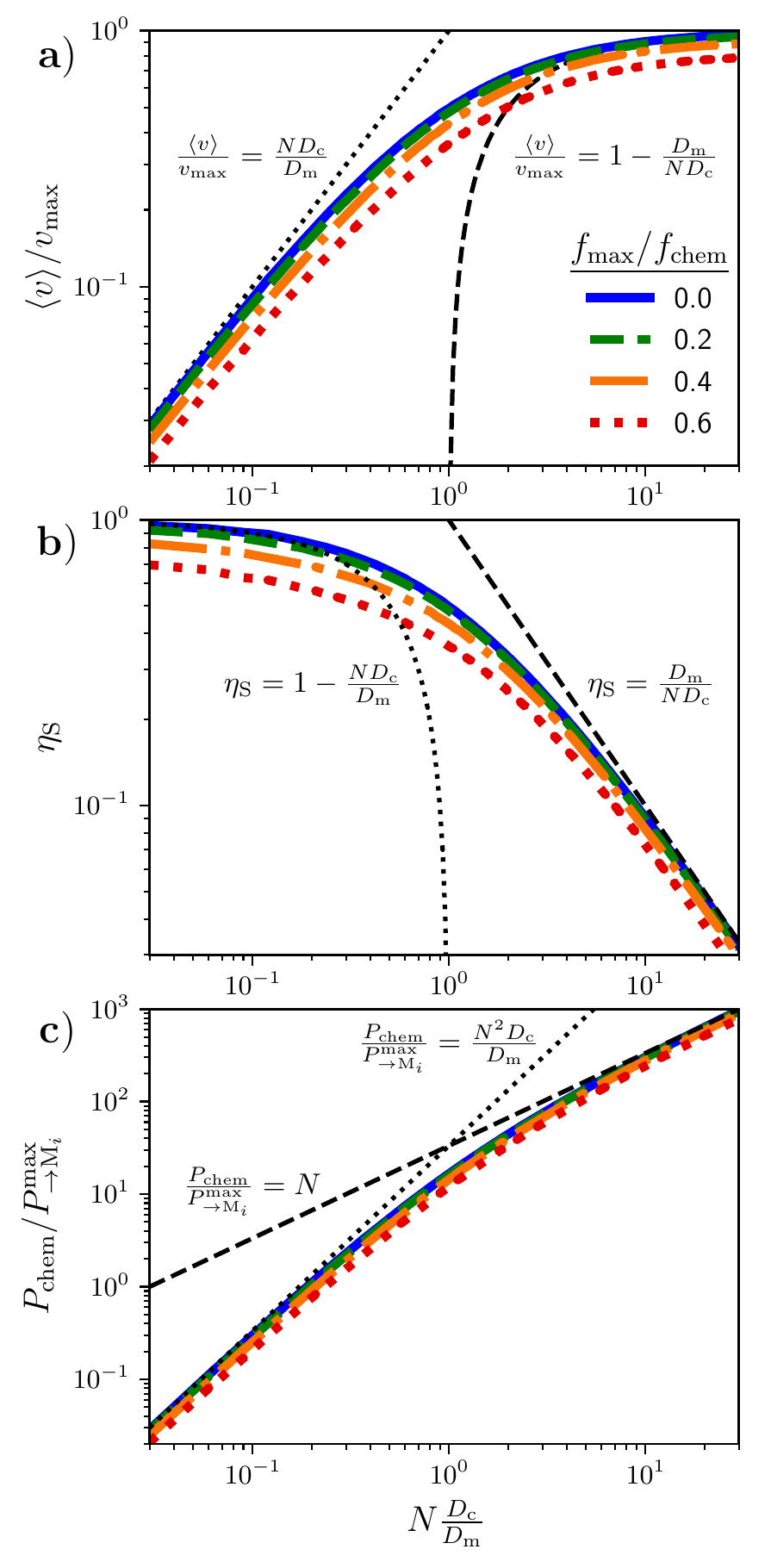}
\caption{\label{fig:simulationscalingfig} 
Scaling of the scaled mean velocity $\langle v\rangle/v_\mathrm{max}$, the Stokes efficiency $\etaS$, and the scaled chemical power consumption $P_\mathrm{chem}/P^\mathrm{max}_{\to\mathrm{M}_i}$, as a function of the motor number $N$ scaled by the diffusivity ratio $\Dm/\Dc$.
We simulate $N\in[1,2,3,\ldots,1000]$ for several different ratios $f_\mathrm{max}/\force$ of the maximum conservative force to the chemical driving force. For $f_\mathrm{max}/\force=0$ (solid blue) we recover our analytically tractable model. Black curves indicate the scaling laws from Table~\ref{tab:table1} in the small-$N$ (dotted) and large-$N$ (dashed) limits. Parameters: $\force = 10$, $\kappa=1$, $\Dc=0.03\Dm$; in dimensionless units chosen so that $D_m=\kappa = \ell = 1$.}
\end{figure}

Figures \ref{fig:simulationscalingfig}a-c shows how three important performance metrics, the mean velocity $\langle v\rangle$, the Stokes efficiency $\etaS$, and the chemical power consumption $P_\mathrm{chem}$, scale with the number $N$ of motors, for varying ratios $f_\mathrm{max}/\force$ between the 
maximum conservative force and the chemical driving force on each motor. In the limit as $f_\mathrm{max}/\force\to0$, this system is identical to the analytically tractable system (described by Eq.~\eqref{multiplemotorlangevin}) that we have focused on in this paper. In the large- and small-$N$ regimes, the scaling of the three performance metrics with $N$ is consistent with the scaling laws from Table~\ref{tab:table1}, shown in black dashed and dotted lines.

\section{Motor binding/unbinding}\label{fluctuating_N}
As discussed in Sec.~\ref{discussion}, we investigate the effects of dynamically changing motor number using a simple stochastic model for motor binding/unbinding based on \cite{klumpp2005cooperative}. The motor number $N$ undergoes a random walk, with rates 
\begin{subequations}
\begin{align}
N\to N+1:\; & \; k_N^+ = k_0^+ (N_\mathrm{max}-N),\\
N\to N-1:\; & \; k_N^- = k_0^- N.
\end{align}
\end{subequations}
Here $k_0^+$ and $k_0^-$ are base rates of binding and unbinding, and $N_\mathrm{max}$ is the maximum number of motors that can bind a given cargo. The distribution $P(N)$ satisfies the master equation
\begin{align}\label{eq:masterN}
&\frac{\partial}{\partial t}P(N) \\
&= k_{N+1}^-P(N+1) + k_{N-1}^+P(N-1) - (k_N^+ + k_N^-)P(N)\nonumber \ ,
\end{align}
with reflecting boundaries at $N=1$ and $N=N_\mathrm{max}$. (We set a reflecting boundary at $N=1$ rather than $N=0$ because we are interested only in the behavior of the system when there are motors attached.) 

The master equation~\eqref{eq:masterN} with these boundary conditions has a time-independent steady-state solution. To simplify the analysis we take the limit $N_\mathrm{max}\to\infty$ and $k_0^+\to 0$ such that $N_\mathrm{max}k_0^+/k_0^-=\lambda$ is fixed. The steady-state solution is then
\begin{equation}\label{eq:N_dist}
P(N) = \frac{\lambda^N}{N!(e^\lambda-1)},
\end{equation}
defined for $N\geq 1$. This is a zero-truncated Poisson distribution \cite{johnson2005univariate}, with mean
\begin{equation}
\langle N\rangle = \frac{\lambda}{1-e^{-\lambda}}.
\end{equation}

We can then use this distribution to calculate mean values of different steady-state properties of the transport system, averaged over $P(N)$. In particular, we estimate the error involved in assuming the system is well described by a constant (rather than fluctuating) number of motors. We use the mean velocity as an example.

For a given average number $\langle N \rangle$ of motors, we compare the mean velocity~\eqref{velocityeq} evaluated at fixed $N=\langle N \rangle$ to the velocity instead averaged over the distribution $P(N)$ with the parameter $\lambda$ chosen so that $\sum_{N=1}^\infty N P(N) = \langle N\rangle$. The error in mean velocity incurred by assuming fixed $N$ is
\begin{equation}\label{eq:error}
\epsilon \equiv \frac{\left\langle v\left(\langle N\rangle_{P(N)}\right)\right\rangle_\mathrm{ss} - \left\langle\langle v (N)\rangle_\mathrm{ss}\right\rangle_{P(N)}}{\left\langle\langle v (N)\rangle_\mathrm{ss}\right\rangle_{P(N)}},
\end{equation}
where $\langle \cdot\rangle_\mathrm{ss}$ denotes an ensemble average over the system dynamics at fixed $N$, and $\langle\cdot\rangle_{P(N)}$ denotes an average over the probability distribution $P(N)$.

\begin{figure}[h]
\includegraphics[width=\columnwidth]{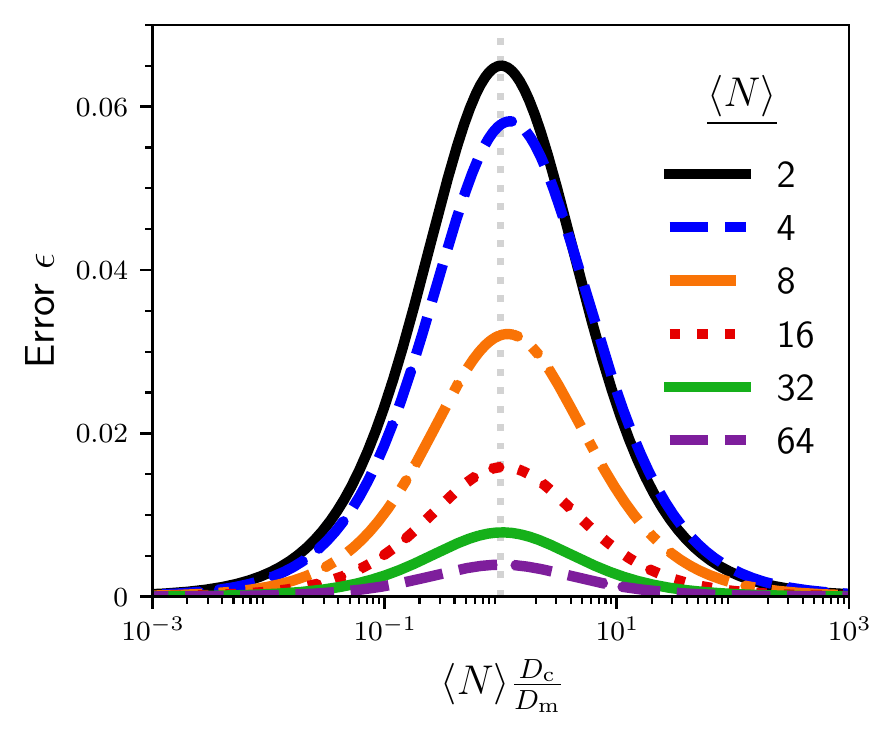}
\caption{\label{fig:fluctuating_N} Error~\eqref{eq:error} in mean velocity between fluctuating (described by Eq.~\eqref{eq:N_dist}) and constant motor number, both with equal mean motor number $\langle N\rangle$, as a function of $\langle N\rangle \Dc/\Dm$.
Vertical gray dotted line indicates $\langle N \rangle \Dc/\Dm = 1$.
}
\end{figure}

Figure \ref{fig:fluctuating_N} shows this error over a range of different values of $\langle N\rangle$ and $\Dc/\Dm$. The error resulting from assuming a fixed number of motors is less than $7\%$, and we find similar magnitudes of error for other quantities. Note that the error is maximized for small $\langle N\rangle$, and for $\langle N\rangle \Dc/\Dm\approx 1$.

\vspace{1ex}
\eject

\end{document}